\documentclass[12pt]{article}

\usepackage{graphicx}

\begin{document}

\title{Predicting solar cycle 24 with a solar dynamo model}

\author{Arnab Rai Choudhuri$^{1}$, Piyali Chatterjee$^{1}$,\\ 
and \\
Jie Jiang$^{2}$\\
$^{1}${\it{Department of Physics, Indian Institute of Science,}} \\
{\it{Bangalore-560012, India}}.\\
$^{2}${\it{National Astronomical Observatories, Chinese Academy of Sciences,}} \\
{\it{Beijing 100012, China.}}}

\date{}
\maketitle

\begin{abstract}
Whether the upcoming cycle 24 of solar activity will be strong or not
is being hotly debated.
The solar cycle is produced by
a complex dynamo mechanism.  We model the last few solar cycles
by `feeding' observational data of the Sun's polar 
magnetic field into our solar
dynamo model.  Our results
fit the observed sunspot numbers of cycles 21-23 extremely
well and predict that cycle~24 will be about 35\% weaker than
cycle~23.
\end{abstract}

Solar activity affects our space environment, thereby influencing
various aspects of human life [{1}].  So it is vitally important to
develop capabilities for predicting strengths of the 11-year cycles
of solar activity. It has been believed for
some time that the Sun's
polar magnetic field at the preceding minimum gives an
indication of the strength of the next solar cycle [{2}].  
The weakness of the present polar field has already
led to predictions that cycle~24 will be the weakest cycle in 100 
years [{ 3, 4}].  Since the solar cycle is produced by a dynamo
mechanism, one would like to make a prediction of cycle~24
from a detailed solar dynamo model also. The only previous
dynamo-based prediction is that cycle~24 will be one of the strongest
cycles [{5}]. Our aim is to generate an independent prediction for
cycle~24 from a different dynamo model by a different methodology.
We identify the Babcock--Leighton
process for poloidal field generation as the main source of
randomness in solar cycles.  A theoretical mean field model
of the solar dynamo produces a poloidal field at the end of
a cycle which would be typical of an `average' solar cycle.
In order to model actual solar cycles, a theoretical mean field 
model of the solar dynamo
has to be `corrected' by feeding actual observational data of
poloidal field.  Since such data are available only from the
mid-1970s, this method can be used to model solar cycles only
from that time. We carry on our calculations by feeding the
DM (Dipole Moment) values of solar 
polar field computed by Svalgaard et al.\ [{3}]
into our already published solar dynamo model [{6, 7}].

Current solar dynamo models combine three basic processes. (i) The
strong toroidal field is produced by the stretching of the poloidal
field by differential rotation in the tachocline at the base of the
convection zone. (ii) The toroidal field generated in the tachocline
gives rise to active regions due to magnetic buoyancy and the decay
of tilted bipolar active regions produces the poloidal field by the 
Babcock--Leighton mechanism.  (iii) The meridional circulation
advects the poloidal field first to high latitudes and then down
to the tachocline.  Two-dimensional mean
field dynamo models based on these three processes were first
constructed about a decade ago [{8, 9}].  We believe that
the processes (i) and (iii) are reasonably smooth and deterministic.
In contrast, the process (ii) involves an element of randomness, which
presumably is the primary cause of solar cycle fluctuations.  Firstly, although
active regions appear in a latitude belt at a certain phase of the
solar cycle, where exactly within this belt the active regions appear
seems random.  Secondly, there is considerable scatter in the tilts
of bipolar active regions around the average given by Joy's law. The action
of the Coriolis force on the rising flux tubes gives rise to Joy's
law [{10}], whereas convective buffeting of 
the flux tubes in the upper layers of the convection zone cause the
scatter of the tilt angles [{11}]. Since the
poloidal field generated from an active region by the Babcock--Leighton 
process depends on the tilt, the scatter in the tilts introduces a
randomness in the poloidal field generation process.  

The poloidal field gets built up during the declining phase of
the cycle and at the minimum, when there are no sunspots, we have
the polar field cumulatively produced from the sunspots during the
previous cycle.
The polar field at the solar minimum produced in a mean field dynamo
model is some kind of `average' polar field during a typical
solar minimum.  The polar field during
a particular solar minimum may be stronger or weaker than this average 
field. We propose the following methodology for modelling the solar
cycles with a mean field dynamo model.  We run the dynamo code in the
usual way from one solar minimum to the next.  Then, at the time of the
minimum, we change the amplitude of the polar field suitably to make
it agree with the observed value of the polar field and run the code
again to the next minimum.  Proceeding in this way, we can correct for
the randomness introduced in the Babcock--Leighton mechanism by using
actual observational data.

Our calculations are based on the solar dynamo code {\em Surya}. This code, 
along with a detailed guide [{12}], is freely
available for use by solar physicists.  Anybody desirous of obtaining
this code may send a request to Arnab Choudhuri through e-mail
(arnab@physics.iisc.ernet.in).
Full details of the two-dimensional 
kinematic dynamo model which is solved by {\em Surya}
are available elsewhere [{7, 12}]. In what 
was referred to as the {\em standard model} 
in Sect.\ 4 of Chatterjee et al.\ [{7}], we change some parameters
to make the period of the solar cycle
equal to 10.6 years (the period in the standard model was 14 years).
The old values and changed values of the parameters are listed in
Table~1. 

\begin{table}
\center
\begin{tabular}{|c|c|c|}
\hline
parameter & Standard Model & This Model \\
\hline 
$v_0$ & $-29$ m s$^{-1}$ &  $-34$ m s$^{-1}$ \\
\hline
$R_p$ & $0.61 R_{\odot}$ & $0.635 R_{\odot}$\\
\hline
$\beta_2$ & $1.8\times10^{-8}$m$^{-1}$ & $1.3\times10^{-8}$ m$^{-1}$ \\
\hline
$r_0$ & $0.1125 R_{\odot}$ & $0.1286 R_{\odot}$ \\
\hline
$d_{tac}$ & $0.05R_{\odot}$ & $0.03R_{\odot}$\\
\hline
\end{tabular}
\caption{The original values of the parameters in the {\em {standard model}} (Sect.\ 4 of Chatterjee et al. [{7}]) along with the changed
values we use now. The first four parameters control the 
amplitude, penetration depth, equatorial return
flow thickness and the position of the inversion layer of
the meridional circulation, respectively. The tachocline width is denoted 
by $d_{tac}$.} 
\end{table}
\def\dma{$\overline{\rm DM}$}

We now discuss how we change the value of the polar field during
successive solar minima to feed the relevant information about the
past cycles into the code.  Reliable data about polar fields
from Wilcox Solar Observatory (WSO) and Mount Wilson Observatory 
(MWO) exist only for the minima at the ends of the cycles~21, 22
and 23.  Additionally, MWO data exist for one previous minimum (at
the end of cycle~20), though the quality of data was not so good at that
time.  Svalgaard et al.\ [{3}] have analyzed these data
carefully and came up with a parameter for the solar minima, which they
call `Dipole Moment (DM)'.  Although we think that this name is
somewhat misleading, we keep using it in this paper.  This DM, which
is a good measure of the polar field during the solar minimum, has
its values for the last 3 solar minima listed in Table~1 of Svalgaard
et al.\ [{3}]. From Fig.~3 of their paper, we estimate
DM for the previous minimum at the end of cycle 20 to be about 
250 $\mu$T, although the data appear noisy.  According to Table~1,
values of DM at the ends of cycles~21, 22 and 23 are respectively
245.1  $\mu$T, 200.8 $\mu$T and 119.3 $\mu$T.  The next question
we have to address is: what value of DM corresponds to the polar
field of an `average' cycle?  This question is not so straightforward
to settle, given the fact that there is a trend of cycle amplitudes
steadily increasing since the Maunder minimum [{13}].  
We tentatively take cycle 23 as an average cycle and
the value of DM before its beginning (which is 200.8 $\mu$T) 
denoted by \dma\ as the average for a typical average solar minimum.
If we divide the DM value of a particular minimum by \dma=200 $\mu$T,
we get a numerical factor which we would call $\gamma$.   The
values of $\gamma$ at the ends of cycles~20, 21, 22 and 23 are
respectively 1.25, 1.23, 1.0 and 0.60.

\def\Am{\overline{A}_{\rm min}}
\def\Rs{R_{\odot}}

The poloidal field in a two-dimensional dynamo problem is described
by a scalar function $A(r, \theta)$.  From a regular run of the dynamo code,
we can find out the value of the amplitude of $A$ at the 
solar minimum, which would correspond to an `average' value for a
typical solar minimum.
Let us call this $\Am$. Suppose we run the dynamo code
till a solar minimum for which we know the value of $\gamma$ from
observational data.  At all grid points above $0.8 \Rs$, we multiply
$A$ by a constant factor such that the amplitude of $A$ becomes
$\gamma \Am$.  We do not make any changes in the values of $A$ below
$0.8 \Rs$.  This ensures that the poloidal field in the upper layers,
which has been created by the Babcock--Leighton mechanism operating
during the last cycle, gets corrected to the observed value, whereas
the poloidal field at the bottom of the convection zone, which may
have been created during the still earlier cycles, is left unchanged.
After changing $A$ above $0.8 \Rs$ in this fashion, we run the code
till the next minimum when this procedure is repeated. 

Since we have values of $\gamma$ at the ends of cycles 20--23, our
procedure for generating a forecast for cycle 24 is now straightforward.
We take a relaxed solution of our dynamo code which has been stopped
at a solar minimum. We identify this minimum as the minimum at the
end of cycle of 20 and change the the values of $A$ above $0.8 \Rs$
in accordance with the value of $\gamma$ (which is 1.25).  Then we
run the code till the next minimum and again change the values of
$A$ above $0.8 \Rs$.  Doing this thrice, we come to the minimum at
the end of the cycle~23.  The next run after this generates the
forecast for cycle~24. It may be noted that the poloidal field lines
become somewhat discontinuous at $r = 0.8 \Rs$ after we change the values
of $A$ above $0.8 \Rs$ in accordance with observational data.  
This discontinuity can be seen in Fig.~1 where
we plot the poloidal field lines at the minimum before cycle~24 just
after updating the values of $A$. However, we find that this discontinuity
gets smoothed out within a time scale of weeks.
 
\def\Rm{R_{\rm max}}

\begin{figure}
\centering{\includegraphics[width=.2\textwidth,height=5.5cm]{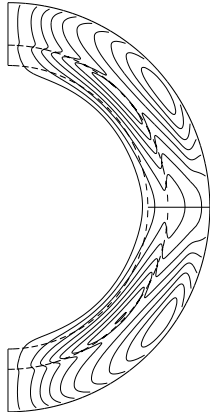}}
\caption{ A snapshot of streamlines of the poloidal field given by constant 
contours of $A r \sin \theta$ just after correcting 
by the DM value for the poloidal field at the 
minimum before cycle 24. The dashed lines correspond to $r =0.7 \Rs$
and $r=0.8 \Rs$.}
\end{figure}

Before presenting results obtained with actual observed values of
DM fed into the code, we present some results obtained by changing
the poloidal field arbitrarily during a solar minimum and then running
the code without any further interruptions. Fig.~2 gives sunspot
number plots obtained by increasing and decreasing 
the poloidal field by 30\% above $0.8 \Rs$ at a solar minimum. 
We find that the next two solar minima
are both affected, after which the memory of the poloidal
field change seems to get lost.    Svalgaard
et al.\ [{3}] suggest a simple relation that the maximum
International Sunspot Number $\Rm$ of cycle~$n$ will be proportional
to the value of DM at the end of cycle~$n-1$, i.e.
$$(\Rm)_n = k ({\rm DM})_{n-1}. \eqno(1)$$
On the basis of our model, we expect a more complicated functional
relationship
$$(\Rm)_n = f [({\rm DM})_{n-1}, ({\rm DM})_{n-2}]. \eqno(2)$$
As we shall see below, the results of our dynamo run for the last few
cycles are in qualitative agreement with what would be expected
from (1).  However, if values of DM during the two previous minima
are widely different, it is in principle possible that our method
based on our dynamo model would generate a forecast for the next
cycle significantly different from what is expected from (1).
\begin{figure}
\centering{\includegraphics[width=.45\textwidth,height=5cm]{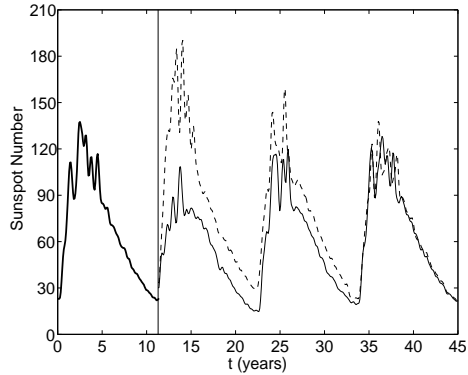}}
\caption{Monthly smoothed sunspot number plots by increasing (dashed line) and 
decreasing (solid line) the poloidal field by 30\% above $0.8 \Rs$
at a solar minimum (indicated by the vertical line), based on our model.
The polar field takes some time to be advected to
the mid-latitudes in the tachocline where a strong toroidal field
is produced during the solar maxima.  In our model, this advection
time is of the order of 10 yr.  After a minimum at the end of cycle~$n$,
the maxima of the next two cycles~$n+1$ and $n+2$ come about 5 yr and
16 yr later respectively.  Since the advection time in our model
is almost an arithmetic mean between these two, the two next maxima
are both affected in our model.  Dikpati \& Gilman [{5}]
point out that the advection time in their model is somewhat longer
(mainly because of the different profiles of meridional circulation
assumed in the two models),
which implies that the poloidal field of cycle~$n$ will primarily
affect the cycle~$n+2$ rather than the cycle~$n+1$.}
\end{figure}
\begin{figure}
\centering{\includegraphics[width=.5\textwidth,height=9cm]{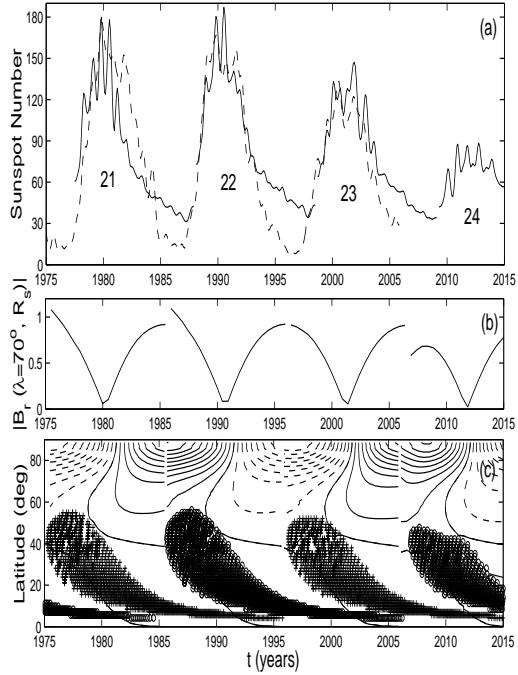}}
\caption{Results for cycles 21--24. (a) The theoretical monthly smoothed sunspot
number (solid line) superposed on the monthly smoothed sunspot numbers from observation (dashed line). (b) A plot of
$B_r$ at the surface at a latitude of 70$^\circ$. (c) The
theoretical butterfly diagram, 
with contours of $B_r$ at the surface
in the time-latitude plot.  In the middle
panel, it is interesting to note that, even though we change $B_r$
abruptly at a minimum, by the next
minimum its value relaxes to values close to what
would be the `average' value for a typical cycle. 
The cycles~21 and 22, which
are of comparable strength, are easy to model as we
see in the top panel, since they follow
solar minima having DM of comparable values.  The cycle~23, which
follows a minimum of low DM, is weak, although we find that the
theoretically calculated cycle is not as weak as the observed cycle.
The cycle~24 is clearly very weak.  The theoretical plot in the
top panel was generated by using \dma = 200 $\mu$T.  The theoretical
plots are found to be qualitatively similar when we take
\dma\ in the range 150--220 $\mu$T. }
\end{figure}

Fig.~3 now presents our results for cycles~21--24 generated by
our methodology. The top panel superposes the
sunspot number generated from our model on the observational data.
The middle panel gives the $B_r$ at a latitude of 70$^\circ$
obtained from the dynamo model, showing the jumps at the solar minima
when we change the poloidal field in accordance with the observed
value of DM. The bottom panel shows the butterfly diagram produced
by our model.   We see in the
top panel that the theoretical plot is in quite good agreement with
the observational data for cycles~21--23.  whereas cycle~24 comes out
as the weakest cycle in a long time.  Since the value of DM during
the minima at the ends of cycles~22 and 23 are lower than the values
of DM in the two preceding minima, the weakness of cycle~24 appears
like a very robust result, which does not change with small changes
in the parameters of the problem such as the chosen value of \dma.
We may point out that the absolute value of the theoretical sunspot number
from our numerical code does not have any particular significance,
since this value changes on changing such things as the grid spacing.
To generate Fig.~3a, we scaled the theoretical sunspot number suitably
to make it fit the observational plot.  

We are now carrying on calculations in which instead of DM we
feed magnetogram data at different latitudes during solar minima
into our model. The results will be presented in a future paper.
While the use of more detailed polar field data may lead to
more realistic predictions, the attractiveness of the scheme
presented in this paper is that it is extremely straightforward
to implement and is probably reasonably reliable, as we have
been able to model cycles 21--23 very well.

Since the dominant processes during the rising phase
of a cycle from a minimum to a maximum are fairly
regular processes like the magnetic field advection and toroidal
field generation by differential rotation, a good knowledge of
magnetic configurations during a minimum should enable a good
theoretical model to predict the next maximum reliably.  On the
other hand, the dominant process in the declining phase of a
cycle is the poloidal field generation by the Babcock--Leighton
process which involves randomness and cannot be predicted in
advance by theoretical models.  In other words, we suggest that
the rising phase of the cycle is predictable (enabling us to predict
the strength of the maximum a few years ahead of time), but
the declining phase is {\em not} predictable.  
Consequently, it may never be
possible to make a realistic prediction of a solar maximum more
than 7--8 years ahead of time, even when we have better theoretical
models and better magnetic data.

Although our forecast is in agreement with physical intuition as 
well as forecasts based on polar field strength [{3, 4}], 
it is completely opposite of
the only other forecast based on a detailed dynamo model [{5}].    
The methodology used by Dikpati \& Gilman [{5}] for feeding the observational
data in the theoretical model differs from ours 
at a fundamental conceptual level.  They use the sunspot area as the
source term for the generation of the poloidal field, whereas
the tacit assumption behind our methodology is that the poloidal
field generation involves randomness and cannot be calculated
deterministically from the past sunspot data. Cycles with many sunspots 
do not necessarily produce strong poloidal fields at the end.  This
is clearly seen in the analysis of Makarov et al.\ [{14}] (see their Fig.~1)
who have used the positions of H$\alpha$ filaments to estimate polar
fields for the better part of a century.  They find that the polar
field during a minimum is correlated with the strength of the next
cycle, but the strength of the cycle has no good correlation with
the polar field produced at its end.  If our
identification of the poloidal field generation by the Babcock--Leighton
process as the main source of randomness in the solar dynamo is correct,
then the methodology of Dikpati \& Gilman [{5}] should in principle not
work, although they claim to `predict' many past cycles correctly. 
Since their forecast for cycle~24 is completely opposite of ours,
it should become apparent in the next 4-5 years as to which forecast
comes closer to truth.

We thank Dibyendu Nandy, Leif Svalgaard and Jingxiu Wang for discussions. We also express our indebtedness to Dibyendu Nandy for his contributions in developing the code {\it{Surya}}. P.\ C.\ acknowledges financial support from Council for Scientific and Industrial Research through grant no. 9/SPM-20/2005-EMR-I. J.\ J. acknowledges financial support from National Basic Research Program of China through grant no. 2006CB806303.

\end{document}